\documentclass{llncs}

\begin{document}

\title{Source-to-source optimising transformations of Prolog programs \\ based on abstract interpretation}

\author{Fran\c{c}ois Gobert \and Baudouin Le Charlier}

\institute{\ \\[-.5cm]
            Universit\'{e} Catholique de Louvain, Belgium, \\
            \email{\{francois.gobert,baudouin.lecharlier\}@uclouvain.be}\\
            This work is supported by the Belgian FNRS fund.}

\maketitle
\vspace{-.6cm}
\begin{abstract}
Making a Prolog program more efficient by transforming its source code, without changing its operational semantics, is not an obvious task. It requires the user to have a clear understanding of how the Prolog compiler works, and in particular, of the effects of `impure' features like the cut. The way a Prolog code is written - e.g., the order of clauses, the order of literals in a clause, the use of cuts or negations - influences its efficiency. Furthermore, different optimisation techniques may be redundant or conflicting when they are applied together, depending on the way a procedure is called - e.g., inserting cuts and enabling indexing. We present an optimiser, based on abstract interpretation, that automatically performs safe code transformations of Prolog procedures in the context of some class of input calls. The method is more effective if procedures are annotated with additional information about modes, types, sharing, number of solutions and the like. Thus the approach is similar to Mercury. It applies to any Prolog program, however.
\\[.1cm]
\end{abstract}


\vspace{-1cm}
\section{Introduction} \label{Introduction}

Programming in Prolog allows us to write so-called {\it multidirectional} procedures, in the sense that the same code of a procedure can be used in more than one way (the arguments being either input data or output results). The undesirable consequence for multidirectionality is inefficiency (in terms of space utilisation and of execution time). This efficiency issue comes from the general execution model of Prolog, which is generally based on the Warren's Abstract Machine (WAM, for short)~\cite{AitKaci91-WAM,Warren83-WAM}. Answer substitutions are computed according to a depth-first search strategy with backtracking, where clauses are executed from top-to-bottom, and literals are executed from left-to-right inside a clause. Given the incompleteness of Prolog, some input-output patterns can loop. Also, Prolog uses a general algorithm for unification, with no restriction on the terms being unified, but most compilers do not perform the occur-check test during the unification. For efficiency reasons, some built-in procedures are not multidirectional (for example, arithmetic and comparison predicates). Negation as failure is sound only if it applies to a ground literal. Due to the incompleteness and unsoundness of Prolog, not every ordering of clauses and literals is operationally correct, and the way a procedure is written greatly influences the search of its solutions, and then, the efficiency.

A solution for optimising multidirectional procedures is to generate specialised code for each particular use of the procedure. In the context of a directionality, one can try to find a more efficient ordering of clauses and literals, such that the program still remains operationally correct for that directionality. In Prolog, we can also try to insert cuts to prune the search tree without removing solutions. This can greatly reduce the size of the search tree, and improves the efficiency. Applying correct code transformations is not obvious and is tricky to be done manually, because it is very error-prone. This paper describes an optimiser based on abstract interpretation, which realizes this task automatically.

To illustrate the interest of specialising code, consider the mutidirectional procedure {\tt efface(X,T,TEff)}, which is the running example of~\cite{Deville90-Book}: {\tt X} is an element of list {\tt T}, and {\tt TEff} is the list {\tt T} without the first occurrence of {\tt X} in {\tt T}.
\begin{center}
\begin{tt}
\begin{tabular}{l}
efface(X,[H|T],[H|TEff]) :- efface(X,T,TEff), not(X=H). \\
efface(X,[X|T],T). \\
\end{tabular}
\end{tt}
\end{center}
This code can be used in several ways: either when every input argument is a ground term; or when inputs {\tt X} and {\tt T} are ground and {\tt TEff} is a variable; or when inputs~{\tt X} and {\tt TEff} are ground and {\tt X} is a variable; or when inputs~{\tt X} and {\tt TEff} are variables and {\tt T} is ground; etc. Now, if we consider only the first directionality, then our optimiser will be able to automatically generate the following specialised code (the optimiser has checked that the procedure is deterministic for this directionality):
\begin{center}
\begin{tt}
\begin{tabular}{l}
efface(X,[X|T],T) :- !. \\
efface(X,[H|T],[H|TEff]) :- efface(X,T,TEff). \\
\end{tabular}
\end{tt}
\end{center}
The clauses are reordered, a cut has been inserted in the first clause, and the negation is removed. For all inputs satisfying the first directionality, the sequences of answer substitutions of the specialised and of the multidirectional codes are identical. Table~\ref{table-efface} compares the execution between the multidirectional and the specialised codes. Several tests have been performed by varying the list-length of the input list {\tt T}. The table shows that the multidirectional code is less efficient than the specialised one in terms of execution time and of used local stack. In particular, the specialised code uses a constant amount of local stack (independently of the size of the input), while we yield a local stack error if we try to execute the multidirectional code with an input list of size 25000. The speedup increases according to the size of the input: for instance, the optimised code spent 3.61 times less execution time for an input list of size 10000.

\begin{table}[h]
\begin{center}
\begin{tabular}{|l|r|r|r|r|r|}
\hline
\multicolumn{1}{|c|}{\bf Input} & \multicolumn{3}{c|}{\bf Execution Time (ms)} & \multicolumn{2}{c|}{\bf Used Local Stack (By)} \\
\cline{2-6}
\multicolumn{1}{|c|}{\bf list}  & \multicolumn{1}{c|}{\bf Multidirectional} & \multicolumn{2}{c|}{\bf Optimized} & \multicolumn{1}{c|}{\bf Multidirectional} & \multicolumn{1}{c|}{\bf Optimized}\\
\multicolumn{1}{|c|}{\bf size}  & \multicolumn{1}{c|}{\bf (ms)} & \multicolumn{1}{c|}{\bf (ms)} & \multicolumn{1}{c|}{\bf (sdup)} & \multicolumn{1}{c|}{\bf (By)} & \multicolumn{1}{c|}{\bf (By)} \\
\hline
100     & 71    & 25    & 2.84  & 8508    & 108\\
1000    & 693   & 225   & 3.09  & 84108   & 108\\
10000   & 8176  & 2264  & 3.61  & 840108  & 108\\
25000   & ERROR & 5633  & -     & ERROR   & 108\\
\hline
\end{tabular}
\end{center}
\caption{Program {\tt efface(X,T,TEFf)} executed 1000 times, when {\tt X} is a ground term, {\tt T} is a ground list, and {\tt TEff} is a variable. The program has been tested on a 1.5~Ghz~Pentium; 1Gb~RAM; Linux~Suse; SWI-Prolog~v~5.4.6~\cite{SwiProlog}. The size to which the local stack is allowed to grow is 2048000 By.} \label{table-efface}
\vspace{-1cm}
\end{table}

Our approach is strongly inspired by~\cite{Deville90-Book}, where a methodology to build correct programs is proposed: starting from a specification and a so-called {\em logic description} of the problem, the methodology constructs operationally correct programs which are not written in the usual style of experienced Prolog programmers: procedures are normalised, with explicit unifications, and are thus inefficient. The author of~\cite{Deville90-Book} then proposes to apply some code transformations, in order to produce more efficient programs (written in the usual style and where cuts are introduced). It is not obvious to ensure the correct application of transformations, nor to choose the best order to apply them. Our optimiser does not require or assume that Prolog programs are written in a specialised syntax. It accepts any kind of Prolog programs. So, the programmer has the liberty to write its program in the style he wants: normalised or not. Our optimiser then automatically specialises the program for some directionality, by choosing a suitable order for applying the transformations, and by ensuring that the transformations are correctly applicable.

The transformations performed by the system are related to partial evaluation: 
literals are evaluated at compile-time, such that some of them can be removed, and some unifications can be simplified. However, we do not unfold procedure calls as it is normally done in partial deduction.

The optimiser is based on an abstract interpretation framework~\cite{Bruynooghe91-PracticalFramework,LeCharlier99-Automatic,LeCharlier02-SequenceBased}, that collects and verifies the semantic information needed for correct application of source-to-source transformations. The operational properties catched and verified by the framework that are useful for the purpose of the optimiser are: cardinality information (including determinacy and conditions for sure success and failure), detection of the exclusivity between clauses, information about the mode, type, sharing, linearity, and size of input/output terms, occur-check freeness, and induction parameters for proving termination.

The rest of the paper is organised as follows. Section~\ref{section2} presents the source-to-source transformations we apply on Prolog programs, and specifies the conditions for applying correctly such transformations. Section~\ref{section3} describes briefly the abstract interpretation framework. Formal specifications are introduced, to allow the user to express the directionality for which the code must be optimised. Section~\ref{section4} illustrates in which order the transformations are applied. Section~\ref{experiment} reports on experimental results, and Section~\ref{rw} presents the related work.


\vspace{-0.2cm}
\section{Source-to-source Transformations} \label{section2}
\vspace{-0.2cm}

Optimisation is carried out by means of transformations based on the operational semantics of Prolog. Section~\ref{PrologSyntax} describes the Prolog programs accepted by the analyser, and the syntax of {\em normalised} programs, on which most code transformations are applied. Section~\ref{SequenceAnswerSubstitutions} introduces the concept of {\em sequence of answer substitutions}, that allows us to describe the sufficient conditions for applying correct source-to-source transformations. Section~\ref{SemanticProperties} defines some semantic properties that characterise sequences of answer substitutions. Such properties are undecidable, but can be safely approximated by our framework. Finally, Section~\ref{TransformationRules} presents the transformation rules.

\vspace{-0.3cm}
\subsection{Prolog Syntax} \label{PrologSyntax}
The optimiser accepts any Prolog program that is ISO conformant, with special features like the cut and the negation, as well as constructs of the form \verb!;! (disjunctions) and \verb!->;! (if then else). The abstract interpretation framework of the optimiser is designed on {\em normalised} programs, and most code transformations apply on such programs.
A normalised procedure $pr$ is a nonempty sequence of clauses $c$. Each normalised clause $c$ has the form $h \verb!:-! g$ where the head is of the form $p(X_1,...,X_n)$, whereas the body $g$ is a possibly empty sequence of normalised literals. A normalised literal is either a built-in of the form $X_{i_1}=X_{i_2}$, a built-in of the form $X_{i_1} = f(X_{i_2},...,X_{i_n})$, a procedure call $p(X_{i_1},...,X_{i_n})$, a cut~$!$, or a negation $not(l)$, where $l$ is a normalised literal. The variables occurring in a literal are all distinct; all clauses of a procedure have exactly the same head; if a clause uses $m$ different variables, these variables are $X_1,...,X_m$. Observe that all Prolog program can be rewritten into equivalent normalised programs.
In the rest of this paper, $P$ denotes the 
given normalised program.

\vspace{-0.3cm}
\subsection{Sequence of Answer Substitutions} \label{SequenceAnswerSubstitutions}
Semantically, a normalised procedure $pr$ can be viewed as a function mapping every input substitution $\theta$ to a sequence of answer substitutions $S$. A substitution~$\theta$ is a finite set of the form $\{X_1/t_1,...,X_n/t_n \}$ where $X_1,...,X_n$ are distinct program variables, and where $t_1,...,t_n$ are terms (variables occurring in terms are {\em standard} variables; the sets of standard and program variables are disjoint). A sequence of answer substitutions $S$ can be either finite $<\theta_1,...,\theta_k>$ ($k\geq0$), or infinite $<\theta_1,...,\theta_k,...>$ ($k \in N$), or incomplete $<\theta_1,...,\theta_k,\bot>$ ($k\geq0$), where the symbol $\bot$ denotes that the procedure loops.
To express this behaviour, we use the notation of~\cite{LeCharlier02-SequenceBased}: $\langle \theta,pr \rangle \mapsto_P S$ for a procedure, $\langle \theta,c \rangle \mapsto_P S$ for a clause, and $\langle \theta,g,c \rangle \mapsto_P S$ for a prefix of a clause.
In the rest of this paper, we assume that each procedure terminates. Thus, we only consider {\em finite} sequences of answer substitutions. Our optimiser is based on the abstract interpretation framework defined in~\cite{LeCharlier99-Automatic}, that uses induction parameters to verify that a procedure terminates. Notice that every procedure of our benchmarks terminates. 
The substitutions of $S$ are denoted by ${\it Subst}(S)$, and the length of $S$ is denoted~by~$|S|$.

\vspace{-0.3cm}
\subsection{Semantic Properties} \label{SemanticProperties}
This section defines some semantic properties that characterise the execution of literals, prefixes of the body of a clause, clauses, procedures, in terms of the length of their sequence of answer substitutions. Such properties are useful to express the sufficient conditions to apply safely the code transformations.

Let $pr$ be a procedure of arity $n$, let $c$ be a clause of the procedure $pr$, let $(g,l)$ be a prefix of the clause $c$, where $g$ is a goal and $l$ is a literal. Let $\theta$ be an input program substitution with domain $\{X_1,...,X_n\}$. Consider the execution of the procedure $\langle \theta, pr\rangle \mapsto_P S_{\it pr}$, the execution of the clause $\langle \theta, c\rangle \mapsto_P S_{\it c}$, the execution of the prefix of the clause $\langle \theta, g, c\rangle \mapsto_P S_{\it g}$. Assume that the literal $l$ is of the form $q(X_{i_1},...,X_{i_r})$. The execution of the literal $l$ after the execution of the goal $g$ can be described for all $\theta' \in {\it Subst}(S_g)$ by $\langle \vartheta_{\theta'}, l \rangle \mapsto_P S_{\theta'}$, where $X_k\vartheta_{\theta'} = X_{i_k}\theta'$ ($1 \leq k \leq r$). We can now define the following properties (the terminology of~\cite{Deville90-Book} is used):

  \begin{itemize}
    \item
      The procedure $pr$, or the clause $c$, or the prefix of a clause $g$, or the literal after the execution of the goal $g$ is {\bf deterministic} w.r.t. $\theta$ iff their sequence of answer substitutions has {\em at most one} computed answer substitution: $|S_{pr}|, |S_{c}|, |S_{g}|, |S_{\theta'}| \in \{0,1\}$, for all $\theta' \in {\it Subst}(S_g)$.
    \item
      The procedure $pr$, or the clause $c$, or the prefix of a clause $g$, or the literal after the execution of the goal $g$ is {\bf fully deterministic} w.r.t. $\theta$ iff their sequence of answer substitutions has {\em one and only one} answer substitution: $|S_{pr}|= |S_{c}|=|S_{g}|=|S_{\theta'}|=1$, for all $\theta' \in {\it Subst}(S_g)$.
    \item
      The procedure $pr$, or the clause $c$, or the prefix of a clause $g$, or the literal after the execution of the goal $g$ {\bf surely succeeds} w.r.t. $\theta$ iff their sequence of answer substitutions has {\em at least one} answer substitution: $|S_{pr}|, |S_{c}|, |S_{g}|,$ $|S_{\theta'}| \geq 1$, for all $\theta' \in {\it Subst}(S_g)$.

    \item
      The literal $l$ is a {\bf test literal} after the execution of the goal $g$ w.r.t. $\theta$ if it is not a cut, and it is deterministic w.r.t. $\theta$, and it does not instantiate any variable. For all $\theta' \in {\it Subst}(S_g)$, $S_\theta'$ is either the empty sequence $<>$ or the sequence $<\vartheta_{\theta'}>$.
    \item
      The two procedures $pr_1$ and $pr_2$ (with the same arity $n$) are {\bf exclusive} w.r.t.~$\theta$ iff either the execution of $pr_1$ fails, or the execution of $pr_2$ fails, or both executions of $pr_1$ and $pr_2$ fail:
      \[
        \left.
        \begin{array}{l}
        \langle \theta, pr_1 \rangle \mapsto_P S_1 \\
        \langle \theta, pr_2 \rangle \mapsto_P S_2 \\
        \end{array}
        \right\}
        \Rightarrow
        S_1 =\; <> \makebox{ or }\; S_2 =\; <>
      \]
      The same definition applies for exclusivity between two clauses of same arity, or between two prefixes of clauses of same arity.
  \end{itemize}
The above properties are undecidable but can be safely approximated by our abstract interpretation framework presented in Section~\ref{section3}.

\vspace{-0.4cm}
\subsection{Transformations Rules} \label{TransformationRules}
\vspace{-0.2cm}

The subsequent transformation rules are adapted from~\cite{Deville90-Book}. Let $pr$ be a normalised procedures whose arity is $n$, and let $\theta$ be a program substitution whose domain is $\{X_1,...,X_n\}$. A rule transforming the procedure $pr$ of the program $P$ into the procedure $pr'$ is {\em correct} w.r.t. $\theta$ if the executions of $pr$ (in the context of the program $P$) and of $pr'$ (in the context of the program $P'$, which is the program $P$ where $pr$ has been replaced by $pr'$) produce the same sequence of answer substitutions. In other words, if $\langle \theta,pr \rangle \mapsto_P S$ and $\langle \theta,pr' \rangle \mapsto_{P'} S'$ then $S=S'$. The conditions given for applying the transformations are expressed using the semantic properties defined in Section~\ref{SemanticProperties}. Such conditions are sufficient but not always necessary. It is assumed that the procedure $pr$ has no side-effects when it is executed with input $\theta$.\\

\vspace{-.4cm}
\noindent{\bf Rule 1: Reorder clauses}
\vspace{-.3cm}
\[
\begin{array}{lcl}
  ... \\
  c_i: p(X_1,...,X_n) &\verb!:-!& g_i. \\
  ... \\
  c_j: p(X_1,...,X_n) &\verb!:-!& g_j. \\
  ... \\
  \hline
  ... \\
  c_j: p(X_1,...,X_n) &\verb!:-!& g_j. \\
  ... \\
  c_i: p(X_1,...,X_n) &\verb!:-!& g_i. \\
  ... \\
\end{array}
\]
\vspace{-.4cm}
where:
\begin{itemize}
\itemsep 0pt
  \item
    for all $k \in \{i,...,j\}$, the clause $c_k$ is deterministic w.r.t. $\theta$;
  \item
    for all $k,l \in \{i,...,j\}: k\not=l$, we have that $c_k$ and $c_l$ are exclusive w.r.t. $\theta$;
  \item
    for all $k \in \{i,...,j\}$, there is no cut in $g_k$.
\end{itemize}

\noindent{\bf Rule 2: Insert green cuts}
\vspace{-.3cm}
\[
\begin{array}{lcl}
  c_1: p(X_1,...,X_n) &\verb!:-!& g_1. \\
  ... \\
  c_k: p(X_1,...,X_n) &\verb!:-!& l_1, ...,l_i, l_{i+1}, ..., l_s. \\
  ... \\
  c_r: p(X_1,...,X_n) &\verb!:-!& g_r. \\
  \hline
  c_1: p(X_1,...,X_n) &\verb!:-!& g_1. \\
  ... \\
  c_k: p(X_1,...,X_n) &\verb!:-!& l_1, ...,l_i, !, l_{i+1}, ..., l_s. \\
  ... \\
  c_r: p(X_1,...,X_n) &\verb!:-!& g_r. \\
\end{array}
\]
\vspace{-.3cm}
where:
\begin{itemize}
\itemsep 0pt
  \item
    the goal $(l_1, ..., l_i)$ is deterministic w.r.t. $\theta$;
  \item
    for all $z \in \{k+1,...,r\}$, the goal $(l_1, ..., l_i)$ and clause $c_z$ are exclusive w.r.t.~$\theta$.
\end{itemize}

\vspace{-.1cm}
\noindent{\bf Rule 3: Eliminate dead code}

\vspace{-.3cm}
\[
\begin{array}{lcl}
  c_1: p(X_1,...,X_n) &\verb!:-!& g_1. \\
  ... \\
  c_k: p(X_1,...,X_n) &\verb!:-!& l_1, ...,l_i, !, l_{i+1}, ..., l_s. \\
  ... \\
  c_r: p(X_1,...,X_n) &\verb!:-!& g_r. \\
  \hline
  c_1: p(X_1,...,X_n) &\verb!:-!& g_1. \\
  ... \\
  c_k: p(X_1,...,X_n) &\verb!:-!& l_1, ...,l_i, !, l_{i+1}, ..., l_s. \\
\end{array}
\]
\vspace{-.4cm}
where:
\begin{itemize}
\itemsep 0pt
  \item
    the goal $(l_1, ..., l_i)$ surely succeeds w.r.t. $\theta$.
\end{itemize}

\vspace{-.2cm}
\noindent{\bf Rule 4: Move backwards cut}
\vspace{-.2cm}
\[
\begin{array}{lcl}
  ... \\
  c_k: p(X_1,...,X_n) &\verb!:-!& l_1, ..., l_{i-1}, !, l_{i}, ..., l_s. \\
  ... \\
  \hline
  ...\\
  c_k: p(X_1,...,X_n) &\verb!:-!& l_1, ..., l_{i-1}, l_{i}, !, ..., l_s. \\
  ... \\
\end{array}
\]
\vspace{-.2cm}
where:
\vspace{-.1cm}
\begin{itemize}
\itemsep 0pt
  \item
    $l_i$ is fully deterministic after the execution of goal $(l_1,...,l_{i-1})$ w.r.t. $\theta$.
\end{itemize}

\vspace{-.2cm}
\noindent{\bf Rule 5: Remove useless test literals}
\vspace{-.2cm}
\[
\begin{array}{lcl}
  c_1: p(X_1,...,X_n) &\verb!:-!& l_1, ..., l_{i-1}, l_i, l_{i+1}, ..., l_s. \\
  ... \\
  c_k: p(X_1,...,X_n) &\verb!:-!& l_1, ..., l_{i-1}, l_i, l_{i+1}, ..., l_s. \\
  ... \\
  c_r: p(X_1,...,X_n) &\verb!:-!& g_r. \\
  \hline
  c_1: p(X_1,...,X_n) &\verb!:-!& l_1, ..., l_{i-1}, l_i, l_{i+1}, ..., l_s. \\
  ... \\
  c_k: p(X_1,...,X_n) &\verb!:-!& l_1, ..., l_{i-1}, l_{i+1}, ..., l_s. \\
  ... \\
  c_r: p(X_1,...,X_n) &\verb!:-!& g_r. \\
\end{array}
\]
\vspace{-.2cm}
where:
\vspace{-.1cm}
\begin{itemize}
\itemsep 0pt
  \item
    $l_i$ is a test literal after the execution of goal $(l_1,...,l_{i-1})$ w.r.t. $\theta$;
  \item
    $l_i$ is deterministic after the execution of goal $(l_1,...,l_{i-1})$ w.r.t. $\theta$;
  \item
    $l_i$ is fully deterministic after the execution of goal $(l_1,...,l_{i-1})$ w.r.t. $\theta$ \\
  {\bf or} \\
    $\exists z \in \{1,...,k-1\}$ such that a cut is surely executed in clause $c_z$ w.r.t $\theta$. \\
\end{itemize}
\vspace{-0.5cm}

The abstract interpretation framework that safely approximates the conditions for applying the above source-to-source transformations is discussed in the next section.

\vspace{-.5cm}
\section{Abstract Interpretation Framework} \label{section3}
\vspace{-.3cm}

This section presents the abstract interpretation framework~\cite{Gobert07-AFADL,LeCharlier99-Automatic} that captures and checks the semantic properties (and other ones) defined in Section~\ref{SemanticProperties}. This semantic information (e.g., about the determinacy, the exclusivity) is needed for ensuring correct application of the code transformations. Section~\ref{AbstractSubstitutionsSequences} describes the two fundamental abstract domains. An {\em abstract substitution} represents a set of program substitutions, and an {\em abstract sequence} represents a set of sequences of answer substitutions. Section~\ref{FormalSpecifications} illustrates the syntax of {\em formal specifications}, that allows the user to write abstract sequences into a convenient syntax. The abstract execution is briefly presented in Section~\ref{AbstractExecution}. Finally, Section~\ref{CheckingSemanticProperties} shows how abstract domains are used to check the semantic properties.

\vspace{-.5cm}
\subsection{Abstract Substitutions and Abstract Sequences} \label{AbstractSubstitutionsSequences}
\vspace{-.15cm}
The domain of {\em abstract substitutions} is an instantiation to modes, types, linearity and possible sharing of the generic abstract domain ${\tt Pat(\Re)}$ described \linebreak {in~\cite{Cortesi00-Combination,GAIA}}. An abstract substitution represents a set of program substitutions of the form $\{X_1/t_1,...,X_n/t_n\}$, where the $X_i$'s are program variables, and the $t_i$'s are terms (variables occurring in terms are standard variables; the set of standard and program variables are disjoint). The set of substitutions represented by~$\beta$ is denoted by ${\it Cc(\beta)}$.  Formally, an abstract substitution $\beta$ is a triple of the form $\langle {\it sv},{\it frm},\alpha\rangle$. The {\em same-value} (${\it sv}$) and {\em frame} (${\it frm}$) components provide information about the structure of terms. Each term described in $\beta$ is represented by an index. The {\it sv} component maps each program variable $X$ to its corresponding index. Hence, the equality ${\it sv}(X)={\it sv}(Y)$ means that variables $X$ and $Y$ are bound to the same term. The {\it frm} component describes the pattern of some indices, by giving their functor name and the indices of their composing subterms. The {\em alpha tuple} $\alpha$ is the generic part of the domain. It provides extra information about all terms and subterms of interest (represented by the indices). In the current analyser, $\alpha$ is of the form $\langle mo,ty,ps,lin,E \rangle$. The $mo$ component~\cite{GAIA} maps each index to its mode (e.g., {\tt gr} (ground), {\tt var} (variable)). The $ty$ component maps each index to its so-called {\em type expression}~\cite{Gobert07-AFADL} (e.g., {\tt list(int)} denotes the set of all lists of integers, {\tt list(any)} denotes the set of all possibly non-instantiated lists). The $ps$ component~\cite{GAIA} is a binary relation over indices, and expresses the possible sharing between two terms. Pairs of indices that do not belong to $ps$ surely do not share a variable. The $lin$ component contains all indices that are surely linear (i.e., they do not contain several occurrences of the same variable). The $E$ component is a set of linear relations between the size of terms (several norms can be combined). The abstract substitution whose concretisation is the empty set is denoted by $\bot$. The greatest lower bound between two abstract substitutions $\beta_1$ and $\beta_2$ is denoted by $\beta_1 \sqcap \beta_2$ and is such that ${\it Cc}(\beta_1) \cap {\it Cc}(\beta_2) = {\it Cc}(\beta_1 \sqcap \beta_2)$.

The domain of {\em abstract sequences} models the operational behaviour of a Prolog procedure. An abstract sequence $B$ describes a set of pairs $\langle \theta,S \rangle$ where $\theta$ is a program substitution and $S$ is the sequence of answer substitutions resulting from executing a procedure (a clause, a goal, etc.) with input substitution $\theta$. An abstract sequence $B$ is a tuple of the form $\langle \beta_{\it in},\beta_{\it ref},\beta_{\it fails},U,\beta_{\it out},E_{\it ref\_out},$ $E_{\it sol}\rangle$ that imposes conditions on the pairs $\langle \theta,S \rangle$. The set of pairs $\langle \theta, S\rangle$ satisfying the conditions imposed by $B$ are denoted by ${\it Cc}(B)$. The {\em input} abstract substitution~$\beta_{\it in}$ describes the class of accepted input calls: $\theta \in {\it Cc}(\beta_{in})$. The {\em refined} abstract substitution $\beta_{\it ref}$ describes the successful input calls, i.e., those that produce at least one solution: $S \not =\; <>$ implies $\theta \in {\it Cc}(\beta_{\it ref})$. The set of abstract substitutions $\beta_{\it fails}$ describes conditions of sure failure: the sequence $S$ is empty if the input substitution satisfies one of the abstract substitution of $\beta_{\it fails}$, i.e.,
if there exists $\beta_f \in \beta_{\it fails}$ such that $\theta\in{\it Cc}(\beta_{f})$ then $S = \; <>$. The {\em untouched} component~$U$ describes the set of input terms that are untouched (non-instantiated) during the execution. The {\em output} abstract substitution $\beta_{\it out}$ describes the substitutions belonging to $S$: for each $\theta'$ in ${\it Subst}(S)$, we have that $\theta' \in {\it Cc}(\beta_{out})$.  The {\em size relations} component $E_{\it ref\_out}$ is a set of linear relations (equations and inequations) between the size of the input/output terms. The {\em cardinality} component $E_{\it sol}$ is a set of linear relations between the number of solutions and the size of the input terms, i.e., {\tt sol=$|S|$} is a solution of the system~$E_{sol}$.

\vspace{-.25cm}
\subsection{Formal Specifications} \label{FormalSpecifications}
\vspace{-.1cm}

Formal specifications describe abstract sequences using a concrete syntax more convenient for a programmer than the mathematical formalism of abstract sequences. A formal specification may contain other information needed for the analyser, like the induction parameter for proving termination. For instance, the following two specifications for the {\tt efface} procedure can be written by the user (and can be checked by the analyser):
\begin{verbatim}
  efface                           efface
    in(X:gr,T:list(gr),TEff:var)     in(X:gr,T:any,TEff:list(gr))
    out(_, _, list(gr))              out(_, list(gr), _)
    srel(TEff_out = T_in-1)          srel(TEff_in = T_out-1)
    sol(sol =< 1)                    sol(sol =< TEff_in+1)
    sexpr(T)                         sexpr(TEff)
\end{verbatim}
We find the different parts of an abstract sequence. The first specification considers the situation where input {\tt X} is a ground term, input {\tt T} is a ground list, and {\tt TEff} is a variable. After success of execution, {\tt TEff} becomes a ground list, whose list-length is the list-length of input {\tt T} minus one. The symbol `\verb!_!' is used when we do not provide refined information about an argument. The execution terminates (the size expression {\tt T} decreases through recursive calls) and is deterministic ({\tt sol=<1}).
In the second specification, input {\tt X} is a ground term, {\tt T} is any term, and {\tt TEff} is a ground list. This execution terminates and is non-deterministic (the number of solutions is between 0 and the list-length of input~{\tt TEff} plus one).

\vspace{-.25cm}
\subsection{Abstract Execution and Annotated Procedures} \label{AbstractExecution}
\vspace{-.1cm}

Abstract execution is performed on normalised procedures (see Section~\ref{PrologSyntax}), because it simplifies the design of abstract operations. The analyser then first translates a general Prolog procedure into an equivalent normalised procedure. The analysis is compositional. The system verifies a procedure against a specification, by assuming that the specifications hold for subproblems. For a given program, it analyses each procedure; for a given procedure, it analyses each clause; for a given clause, it analyses each atom. If an atom in the body of a clause is a procedure call, the analyser looks at the given specifications to infer information about its execution. The analyser succeeds if, for each procedure and each specification describing this procedure, the analysis of the procedure yields results that are covered by the considered specification.

As a result of the analysis, the procedure is annotated with abstract sequences at each program point. The annotation of the clause $c ::= p(X_1,...,X_n) \verb!:-! l_1,...,l_s.$ in the context of an abstract sequence $B_{p} = \langle \beta_{\it in}^{\it p}, ... \rangle$ is of the form:
\[
    (\beta_{\it in}^{\it p}) p(X_1,...,X_n) \verb!:-! (B_0) l_1, (B_1) ..., l_s (B_s) . (B_{\it c})
\]
The analyser certifies that every abstract sequence at a program point safely approximates the sequence of answer substitutions computed until that point. Let $\theta$ be an input program substitution in ${\it Cc}(\beta_{\it in}^{\it p})$. For each program point~$i$ ($0\leq i \leq s$), the concrete execution of $l_1,...,l_i$ is approximated by $B_i$, i.e., \linebreak $\langle \theta, (l_1,...,l_i) , c\rangle \mapsto_P S_i$ implies $\langle \theta, S_i \rangle \in {\it Cc}(B_i)$. Similarly, the whole clause execution is approximated by $B_{\it c}$, i.e., $\langle \theta, c\rangle \mapsto_P S$ implies $\langle \theta, S \rangle \in {\it Cc}(B_c)$.
The annotation of a procedure in the context of an abstract sequence $B_p$ is the sequence of its annotated clauses.

\vspace{-.1cm}
\subsection{Checking Semantic Properties} \label{CheckingSemanticProperties}
\vspace{-.1cm}

This section explains how the analyser can check the semantic properties of Section~\ref{SemanticProperties} that are useful for applying the code transformations. The components of the abstract sequences provide constraints about the length of the computed answer substitutions.

Let ${\it B = \langle\beta_{in},\beta_{ref},\beta_{fails},U,\beta_{out},E_{ref\_out},E_{sol}\rangle}$ be the abstract sequence approximating the execution of a procedure, or of a clause, or of a prefix in a clause, or a literal $l$ after the execution of a goal. Let ${\it B_i = \langle\beta_{in},\beta_{ref}^i,\beta_{fails}^i,U^i,}$ ${\it \beta_{out}^i,}$ ${\it E_{ref\_out}^i,E_{sol}^i\rangle}$ be the abstract sequence modelling the execution of the clause~$c_i$, or the execution of some prefix of the clause $c_i$ ($1\leq i \leq 2$), where $c_1$ and $c_2$ have the same name and arity.
  \begin{itemize}
    \item
      If ${\bf deterministic}(B)$ returns true then $\langle \theta,S \rangle \in {\it Cc}(B)$ implies $|S| \leq 1$.
      The value of ${\bf deterministic}(B)$ is set to true if {\tt sol=<1} is a solution of $E_{\it sol}$.

    \item
      If ${\bf fully\_deterministic}(B)$ returns true then $\langle \theta,S \rangle \in {\it Cc}(B)$ implies ${|S|=1}$.
      The value of ${\bf fully\_deterministic}(B)$ is set to true if $\beta_{\it in}=\beta_{\it ref}$, and ${\beta_{\it fails}=\emptyset}$, and {\tt sol=1} is a solution of $E_{\it sol}$.
      
    \item
      If ${\bf test\_literal}(B)$ returns true then $\langle \theta,S \rangle \in {\it Cc}(B)$ implies $S = \; <>$ or {$S = \; <\theta>$}.
      The value of ${\bf test\_literal}(B)$ is set to true if {\tt sol=<1} is a solution of $E_{\it sol}$ and if the {\it untouched} component $U$ contains all the indices of $\beta_{\it ref}$.

    \item
      Let $\langle \theta,S_1 \rangle \in {\it Cc}(B_1)$ and $\langle \theta,S_2 \rangle \in {\it Cc}(B_2)$.
      If ${\bf exclusive}(B_1,B_2)$ returns true then $S_1 = \; <>$ or $S_2 = \; <>$. 
      The value of ${\bf exclusive}(B_1,B_2)$ is set to true if one of the three following conditions holds:
      \begin{itemize}
        \item the {\em refined} components $\beta_{\it ref}^1$ and $\beta_{\it ref}^2$ are incompatible: $\beta_{\it ref}^1 \sqcap \beta_{\it ref}^2 = \bot$
        \item or $\exists \beta_{f}^1 \in \beta_{\it fails}^1 : (\beta_{\it ref}^1 \sqcap \beta_{\it ref}^2) \leq \beta_f^1$
        \item or $\exists \beta_{f}^2 \in \beta_{\it fails}^2 : (\beta_{\it ref}^1 \sqcap \beta_{\it ref}^2) \leq \beta_f^2$
      \end{itemize}
  \end{itemize}

The accuracy and the cooperation between the abstract domains allow the analyser to detect automatically whether the conditions are satisfied for applying code transformations. The next section discusses the transformation strategy realized by the optimiser.

\vspace{-.2cm}
\section{A Strategy to Generate Specialised Code} \label{section4}
\vspace{-.2cm}

In the derivation of a procedure, there is often more than one possible sequence of transformations, resulting in different procedures. Some heuristics must then be included whithin the automation of the derivation of procedures to find permutations leading to efficient procedures for a given specification. For instance, the following heuristics are suggested in~\cite{Deville90-Book}:
\begin{itemize}
  \item
    {\it Choose tail recursive permutations.} Last call optimisation is implemented in the WAM~\cite{Warren83-WAM}, such that the environment is deallocated {\it before} executing the last call. The efficiency gain is substantial if the last literal is a recursive call.
  \item
    {\it Choose permutations of the literals with the longest deterministic prefix.} This choice prevents useless computed answer substitutions by prefixes of the literals. Multiple answer substitutions are only generated by the suffixes.
  \item
    {\it Choose permutations of the literals that support the introduction of cuts and the removal of useless literals.}
  \item
    {\it Choose permutations of the literals such that the unifications are at the beginning.} This is useful to instantiate the clause heads and to suppress these equality literals.
\end{itemize}

The combination of transformation rules performed by the optimiser is guided by the above heuristics, and is illustrated on the example {\tt efface}, whose initial multidirectional source code is:
\vspace{-.2cm}
\small
\begin{center}
\begin{tt}
\begin{tabular}{l}
efface(X,[H|T],[H|TEff]) :- efface(X,T,TEff), not(X=H). \\
efface(X,[X|T],T). \\
\end{tabular}
\end{tt}
\end{center}
\normalsize
\vspace{-.2cm}
The following directionality is considered (expressed into a formal specification):
\vspace{-.5cm}
\small
\begin{verbatim}
         efface
           in(X:gr, T:list(gr), TEff:any)
           sol(sol =< 1)
\end{verbatim}
\normalsize

{\bf STEP A: Syntactic normalisation.}
Most code transformations apply on normalised programs, and the abstract execution itself is defined on normalised programs, because it facilitates the analysis. Thus, the first step of the optimiser consists of translating the original procedure into an equivalent normalised code:
\small
\begin{center}
\begin{tt}
\begin{tabular}{lcl}
efface(X1,X2,X3) &:-& X2=[X4|X5], X3=[X4|X6], efface(X1,X5,X6), not(X1=X4). \\
efface(X1,X2,X3) &:-& X2=[X1|X3]. \\
\end{tabular}
\end{tt}
\end{center}
\normalsize

{\bf STEP B: Code annotation.}
In this step, the information needed to apply the code transformations is captured by the checker. Every clause of the normalised procedure is annotated with abstract sequences at each program point.

{\bf STEP C: Clause reordering.}
If the order of solutions is not modified, some clause reordering is achieved ({\bf Rule~1}): a clause containing a literal that is a good candidate to be removed after the possible introduction of cuts, is placed at the bottom of the procedure. In the example {\tt efface} and for the considered directionality, the following code is generated (clause reordering can be realized because the procedure is deterministic):
\small
\begin{center}
\begin{tt}
\begin{tabular}{lcl}
efface(X1,X2,X3) &:-& X2=[X1|X3]. \\
efface(X1,X2,X3) &:-& X2=[X4|X5], X3=[X4|X6], efface(X1,X5,X6), not(X1=X4). \\
\end{tabular}
\end{tt}
\end{center}
\normalsize

{\bf STEP D: Semantic normalisation.}
In order to insert a cut at the {\em leftmost} position in a clause, it may be useful to decompose a unification that may fail into equivalent but simpler unifications. It may then happen that a cut will be placed between such unifications, instead of after the (unique) global unification. This step is called {\it semantic normalisation} to distinguish it with the {\it syntactic normalisation} performed at the step~A: the optimiser uses the semantic information available at each program point about the structure, the mode, the type, the sharing and the linearity of terms, as well as the sure success of the execution of a unification. The following code is generated for {\tt efface} (the last clause is not normalised semantically because no cut will be inserted there):
\small
\begin{center}
\begin{tt}
\begin{tabular}{lcl}
  efface(X1,X2,X3) &:-& X2=[X4|X5], X4=X1, X5=X3. \\
  efface(X1,X2,X3) &:-& X2=[X4|X5], X3=[X4|X6], efface(X1,X5,X6), not(X1=X4). \\
\end{tabular}
\end{tt}
\end{center}
\normalsize
In the first clause, the unification {\tt X2=[X1|X3]} has been decomposed in three elementary unifications: the initial unification succeeds if {\tt X2} is a non-empty list (i.e., {\tt X2=[X4|X5]}), and if the first element of {\tt X2} is {\tt X1} (i.e., {\tt X4=X1}), and if the tail of {\tt X2} is {\tt X3} (i.e., {\tt X5=X3}).

{\bf STEP E: Leftmost cut insertion with dead code elimination.}
The objective of this step is to insert {\it leftmost} green cuts in every clause, except in the last clause ({\bf Rule~2}). In our example, a green cut is introduced in the first clause, at the program point where it is exclusive with the second clause:
\small
\begin{center}
\begin{tt}
\begin{tabular}{lcl}
  efface(X1,X2,X3) &:-& X2=[X4|X5], X4=X1, !, X5=X3. \\
  efface(X1,X2,X3) &:-& X2=[X4|X5], X3=[X4|X6], efface(X1,X5,X6), not(X1=X4). \\
\end{tabular}
\end{tt}
\end{center}
\normalsize
The cut has been inserted before the last unification {\tt X5=X3}. If we had not performed the previous step, then the cut would have been placed at the end of the first clause. 
When inserting a cut, some successive clauses may become useless, such that they can be removed ({\bf Rule~3}).

{\bf STEP F: Move cuts backwards.}
The objective of this step is to obtain the longest deterministic prefix before executing a cut in a clause.
While making sure that the procedure still terminates and that the subcalls are still correctly moded and typed, some literals are reordered inside the clauses. 
An inserted cut is moved backwards by passing literals that surely succeed before the cut ({\bf Rule~4}). In our example, the cut cannot be moved backwards, because the unification {\tt X5=X3} does not surely succeed (at that point, {\tt X5} is a ground list and~{\tt X3} is any term).

{\bf STEP G: Removing useless literals.}
The analyser is able to capture the input conditions for which a cut is {\em surely} executed. This information is used to refine the input conditions of the successive clauses. This allows the optimiser to remove some literals which become useless ({\bf Rule~5}) (e.g., negation, test predicates, arithmetic built-ins). In our example, the cut is surely executed when the first element of input list {\tt X2} is the input {\tt X1}. The second clause is thus surely not executed for that input. In particular, the negation surely succeeds and can therefore be suppressed safely:
\small
\begin{center}
\begin{tt}
\begin{tabular}{lcl}
  efface(X1,X2,X3) &:-& X2=[X4|X5], X4=X1, !, X5=X3. \\
  efface(X1,X2,X3) &:-& X2=[X4|X5], X3=[X4|X6], efface(X1,X5,X6). \\
\end{tabular}
\end{tt}
\end{center}
\normalsize

{\bf STEP H: Semantic denormalisation.}
The code generated until this step is still normalised (with possibly added cuts). Thus, it remains inefficient. The last step consists of applying the reverse transformation. The {\em semantic denormalisation} uses the information captured by the analyser to suppress explicit unifications, or to replace them by simpler ones, and to place them implicitly in the head of clauses. The specialised code for {\tt efface} is thus finally generated:
\small
\begin{center}
\begin{tt}
\begin{tabular}{lcl}
  efface(X1,[X1|X2],X3)      &:-& !, X2=X3. \\
  efface(X1,[X4|X2],[X4|X3]) &:-& efface(X1,X2,X3). \\
\end{tabular}
\end{tt}
\end{center}
\normalsize

{\bf Remark.}
It may happen that the initial source code already contains some cuts. In such situation, the optimiser first removes every green cut, such that only the necessary {\em leftmost} cuts will be introduced during the step~E.
\vspace{-0.5cm}
\section{Experimental Evaluation} \label{experiment}
\vspace{-0.2cm}

Table~\ref{table-efface-gga} compares the execution between the original and the generated versions of {\tt efface(X,T,TEff)} presented in the previous section. Several tests have been performed by considering executions that succeed and that fail, and by varying the list-length of the input list {\tt T}. The table shows that the multidirectional code is less efficient than the specialised one in terms of execution time and of used local stack. In particular, the specialised code uses a constant amount of local stack (independently of the size of the input), while we yield a local stack error if we try to execute the multidirectional code with an input list of size 25000.

\begin{table}[h]
\vspace{-0.5cm}
\begin{center}
\begin{tabular}{|c|l|r|r|r|r|r|}
\hline
\multicolumn{1}{|c|}{\bf Success} &\multicolumn{1}{|c|}{\bf Input} & \multicolumn{3}{c|}{\bf Execution Time (ms)} & \multicolumn{2}{c|}{\bf Used Local Stack (By)} \\
\cline{3-7}
\multicolumn{1}{|c|}{\bf of}&\multicolumn{1}{|c|}{\bf list}  & \multicolumn{1}{c|}{\bf Multidirectional} & \multicolumn{2}{c|}{\bf Optimized} & \multicolumn{1}{c|}{\bf Multidirectional} & \multicolumn{1}{c|}{\bf Optimized}\\
\multicolumn{1}{|c|}{\bf execution} & \multicolumn{1}{|c|}{\bf size}  & \multicolumn{1}{c|}{\bf (ms)} & \multicolumn{1}{c|}{\bf (ms)} & \multicolumn{1}{c|}{\bf (sdup)} &  \multicolumn{1}{c|}{\bf (By)} & \multicolumn{1}{c|}{\bf (By)} \\
\hline
no & 100     & 41    & 23    & 1.83  & 8508    & 156\\
no & 1000    & 380   & 194   & 1.96  & 84108   & 156\\
no & 10000   & 4268  & 1893  & 2.25  & 840108  & 156\\
no & 25000   & ERROR & 4715  & -     & ERROR   & 156\\
\hline
yes & 100     & 115   & 26    & 4.39  & 8508    & 156\\
yes & 1000    & 689   & 221   & 3.11  & 84108   & 156\\
yes & 10000   & 7334  & 2207  & 3.32  & 840108  & 156\\
yes & 25000   & ERROR & 5527  & -     & ERROR   & 156\\
\hline
\end{tabular}
\end{center}
\caption{Program {\tt efface(X,T,TEFf)} executed 1000 times, when {\tt X} is a ground term,~{\tt T} is a ground list, and {\tt TEff} is any term. Several tests are performed, depending on whether the execution fails or succeeds, and depending on the list-length of~{\tt T}. The program has been tested on a 1.5~Ghz~Pentium; 1Gb~RAM; Linux~Suse; SWI-Prolog~v~5.4.6~\cite{SwiProlog}. The size to which the local stack is allowed to grow is 2048000~By.} \label{table-efface-gga}
\vspace{-1cm}
\end{table}

The optimiser has been tested on some classical programs, borrowed \linebreak from~\cite{Apt97-Book,Deville90-Book,SterlingShapiro94-ArtOfProlog} and from the Internet. The source programs, the formal specifications, the generated specialised codes, and the efficiency tests are available at \linebreak {\it http://www.info.ucl.ac.be/$\sim$gobert}. The tests have been realized on a 1.5~Ghz~Pentium, 1Gb~RAM, Linux~Suse, with SWI-Prolog~v~5.4.6~\cite{SwiProlog}.

{\bf Tests on Execution Time.} Table~\ref{table-time} reports on execution time speedup, defined as the ratio between the execution time spent for the source program and for the specialised program. A speedup greater than (resp. less than) one means that the specialised code is more (resp. less) efficient than the source code. We consider 59 procedures and 112 specifications (some procedures have several specifications, because they are multidirectional). The benchmark {\it all} is composed of 173 efficiency tests (there is at least one efficiency test for each specification of a procedure). The benchmark {\it det} is a subset of the benchmark {\it all}. It contains only the efficiency tests for the deterministic procedures. The benchmark {\it ss} contains the efficiency tests for the procedures that surely succeed. The benchmark {\it det+ss} contains the efficiency tests for the determinitic procedures that surely succeed. The mean speedup ranges from 1.42 to 1.68. The maximal speedup is~8.54 and the minimum speedup is~0.59.

\begin{table}[h]
\vspace{-0.5cm}
\begin{center}
\begin{tabular}{|ll|c|c|c|}
\hline
\multicolumn{2}{|c|}{{\bf Efficiency}} & \multicolumn{3}{c|}{{\bf Execution Time (Sdup)}} \\
\cline{3-5}
\multicolumn{2}{|c|}{{\bf tests}}      & \multicolumn{1}{c|}{\bf $\;\;$Mean$\;\;$} & \multicolumn{1}{c|}{\bf $\;\;$Max$\;\;$} & \multicolumn{1}{c|}{\bf Min} \\
\hline
173 & {\it all}     & 1.42  & 8.54 & 0.59 \\
125 & {\it det}     & 1.45  & 8.54 & 0.59 \\
45  & {\it ss}      & 1.6   & 8.54 & 0.8  \\
38  & {\it det+ss}  & 1.68  & 8.54 & 0.8  \\
\hline
\end{tabular}
\end{center}
\caption{Execution time speedup between source codes and specialized codes generated by the optimiser: 173 efficiency tests distributed out of 59 predicates and 112 formal specifications. From the 173 tests, we obtain a speed up for 112 tests, and a speed down for 61 tests.} \label{table-time}
\vspace{-1cm}
\end{table}

{\bf Tests on Space Utilisation.} Table~\ref{table-space} reports on local stack utilisation. 57 generated procedures are considered. The maximal amount of local stack used during the execution of the generated code is either reduced (for 28 procedures), or identical (17 procedures), or increased (12 procedures) w.r.t. the maximal amount of local stack used during the execution of the source code.

\begin{table}[h]
\vspace{-0.5cm}
\begin{center}
\begin{tabular}{|c||c|c|c|}
\hline

{\bf Generated}   & {\bf Used local stack}  & {\bf Used local stack}  & {\bf Used local stack}  \\
{\bf procedures}  & {\bf is reduced wrt}    & {\bf is identical wrt}  & {\bf is increased wrt}   \\
                  & {\bf source code}       & {\bf source code}       & {\bf source code}  \\
\hline
57 & 28 & 17 & 12 \\
\hline
\end{tabular}
\end{center}
\caption{Comparisons between source \& generated codes in terms {of~space~utilization.}} \label{table-space}
\vspace{-1cm}
\end{table}

{\bf Accuracy of the Tests.} The results reported on Table~\ref{table-time} and Table~\ref{table-space} depend on the choice of the efficiency tests. It is impossible to perform tests that include every situation in the context of a directionality. We have tried to make sufficiently general tests for each directionality. The efficiency results depend also on the way the source code is written. For all the benchmark reported in the tables, the original program was written in the usual Prolog style (i.e., not normalised). The optimiser can take as input programs that are normalised, like the ones that are derived in the methodology~\cite{Deville90-Book}. If we perform the same efficiency tests with normalised programs as initial source code, then we obtain a mean speedup of~3. This shows the utility of the optimiser.

{\bf Generated code may sometimes be less efficient.} In general, the optimiser generates code that is more efficient than the source code, in terms of execution time and of local stack utilisation. But the tables report that a generated code can sometimes be less efficient than the original code. For instance, this occurs with the {\tt append(L1,L2,L3)} procedure which concatenates two lists (we consider the usual directionality where inputs {\tt L1} and {\tt L2} are ground lists and {\tt L3} is a variable). The initial source code is:
\vspace{-.2cm}
\small
\begin{center}
\begin{tt}
\begin{tabular}{l}
append([],L,L). \\
append([H|L1],L2,[H|L3]) :- append(L1,L2,L3). \\
\end{tabular}
\end{tt}
\end{center}
\normalsize
\vspace{-.2cm}
In SWI-Prolog, indexing is enabled by default on the first argument. Thus, in the context of the considered directionality, no choice point is created on the local stack. Indeed, the right clause is directly selected according to the principal functor of the first input list, which is either an empty or a non-empty list. Furthermore, the second clause is a {\em chain rule} (i.e., there is only one atom in the body), such that no environment frame is allocated on the local stack. Therefore, this source code uses a constant amount of local stack, and executes very quickly.

The optimiser does not take into account the indexing technique in its strategy for applying correct code transformations, and the following code will be generated:
\vspace{-.3cm}
\small
\begin{center}
\begin{tt}
\begin{tabular}{l}
append([H|L1],L2,[H|L3]) :- append(L1,L2,L3), !. \\
append(\_,L,L). \\
\end{tabular}
\end{tt}
\end{center}
\normalsize
\vspace{-.1cm}
where the two clauses are reordered, a cut is inserted after the recursive call, and the constant {\tt []} in the second clause is removed. This code is less efficient than the initial source code. Indexing has no effect because the right clause cannot be selected according to the principal functor of the input list {\tt L1}. Thus, Prolog creates a choice point on the local stack each time the first clause is called (and it is often executed because it is the recursive call). The cut is a {\it deep} cut (i.e., it is not located just after the symbol {\tt :-}). It occurs after the recursive call, such that the first clause is no more a {\em chain rule}, and an environment frame must be allocated on the local stack each time we execute the first clause. The amount of local stack used during execution is increasing through recursive calls, to the contrary of the initial source code, which uses a constant amount of local stack.

\vspace{-0.3cm}
\section{Related Work} \label{rw}
\vspace{-0.2cm}

The Mercury programming language~\cite{Somogyi96-Mercury} is associated with a whole range of analysis tools for optimisation purposes. In Mercury also, the programmer has to annotate the program with information about modes, types, success and determinacy. A main difference is that not all logic programs are accepted by Mercury (only limited forms of unification are allowed). There are restrictions on the form of the programs and queries in order to generate more efficient code. Mercury is not based on the Warren's Abstract Machine, but has specialised - more efficient - algorithms, depending on the determinacy information. So our approach is more appealing to programmers who are willing to keep the full power of Prolog. Unlike Mercury, we do not change the usual execution model of Prolog based on the WAM~\cite{AitKaci91-WAM,Warren83-WAM}: our optimiser performs some transformations at the Prolog code level. Actually, our optimiser should be able to generate low-level code. But performing source-to-source transformations is more portable (and easier to explain) than generating specialised WAM's instructions. Most Mercury annotations can be translated into our formal specification language. A current limitation to our language is that, for instance, we cannot express that an input list contains only free distinct variables (this can be expressed in Mercury). On the other hand, more general directionalities can be described in our language. For instance, we can express that some argument is a list possibly non-instantiated and possibly non-linear, and that two terms possibly share a variable (this cannot be expressed in Mercury), like in the following formal specification for {\tt append}:
\vspace{-.1cm}
\small
\begin{verbatim}
      append
        in(L1:list(any), L2:list(any), L3:any)
        out(_, _, list(any))
        sol(sol =< 1)
\end{verbatim}
\normalsize
\vspace{-.1cm}

The Ciao preprocessor CiaoPP~\cite{Hermenegildo05-IntegratedProgramDebuggingVerificationOptimization} is a powerful static analyser based on abstract interpretation, which features many analyses similar to ours and other ones. The system can infer and/or check properties like regular types, modes, sharing, non-failure and determinacy, bounds on computational cost, bounds on sizes of terms in the program, and termination. In that system, procedures can be optionally annotated by assertions, which partially corresponds to specifications of our system. The system can perform automatic optimisations such as source-to-source transformation, specialisation, partial evaluation of programs, program parallelisation. Some transformations like cut insertion and semantic denormalisation are not performed in CiaoPP.

The authors of~\cite{Debray90-BanishingCut} consider how most of the common uses of cut can be eliminated from Prolog source programs, by relying on static analysis to generate them at compile time. Static analysis techniques are used to detect situations where to place cuts. In our approach, the insertion of cuts is only one part of the optimisation process: several source-to-source transformations are applied to find the best place where to place the cut (e.g., clause and literal reordering, semantic normalisation, etc.), to remove literals becoming useless due to the execution of cut in previous clauses, and to perform some partial evaluation.
\vspace{-0.5cm}
\section{Conclusion and Future Work} \label{conclusion}
\vspace{-0.3cm}

We have presented an optimiser based on abstract interpretation which attempts to make a Prolog program more efficient by transforming its source code without changing its operational semantics. The tool automatically performs safe code transformations of any Prolog program in the context of some class of input calls (described in formal specifications). Preliminary experimental tests of the optimiser show encouraging results, since the specialised codes are, at the average, 1.42 time more efficient in terms of execution time consumption and of local stack utilisation. The optimiser can be improved: we plan to find better heuristics for specialising the code, based on the knowledge of the WAM~\cite{AitKaci91-WAM,Warren83-WAM}, in order to take into account the indexing technique, and the influence on the efficiency of adding or not {\it deep} cuts.

\vspace{-0.5cm}

\bibliography{wlpe07.bbl}
\bibliographystyle{plain}


\end{document}